\documentclass[pre,aps,twocolumn, superscriptaddress]{revtex4}
\usepackage{amsmath}
\usepackage{epsfig}
\usepackage{amssymb}
\usepackage{color}

\begin{document}

\title{Isothermal Langevin dynamics in systems with power-law
  spatially-dependent friction}

\author{Shaked Regev} 
\affiliation{Department of Biomedical
Engineering, Ben-Gurion University of the Negev, Be'er Sheva 85105,
Israel} 

\author{Niels Gr\o nbech-Jensen} 
\affiliation{Department of Mechanical
and Aerospace Engineering, University of California, Davis, California
95616, USA} 
\affiliation{Department of Mathematics, University of
California, Davis, California 95616, USA}

\author{Oded Farago} 
\affiliation{Department of Biomedical
Engineering, Ben-Gurion University of the Negev, Be'er Sheva 85105,
Israel} 
\affiliation{Ilse Katz Institute for Nanoscale Science and
Technology, Ben-Gurion University of the Negev, Be'er Sheva 85105,
Israel}

\begin{abstract}

We study the dynamics of Brownian particles in a heterogeneous
one-dimensional medium with a spatially-dependent diffusion
coefficient of the form $D(x)\sim |x|^c$, at constant temperature. The
particle's probability distribution function (PDF) is calculated both
analytically, by solving Fick's diffusion equation, and from numerical
simulations of the underdamped Langevin equation. At large times, the
PDFs calculated by both approaches yield identical results,
corresponding to subdiffusion for $c<0$, and superdiffusion for
$0<c<1$. For $c>1$, the diffusion equation predicts that the particles
accelerate. Here, we show that this phenomenon, previously considered
in several works as an illustration for the possible dramatic effects
of spatially-dependent thermal noise, is unphysical. We argue that in
an isothermal medium, the motion cannot exceed the ballistic limit
($\left\langle x^2\right\rangle \sim t^2$). The ballistic limit is
reached when the friction coefficient drops sufficiently fast at large
distances from the origin, and is correctly captured by Langevin's
equation.

\end{abstract} 

\maketitle

\section {Introduction}
\label{sec:intro}

Brownian motion was first observed by the botanist Robert Brown while
examining, under a microscope, the motion of pollen grains and
noticing their random jitter in water \cite{brown}. Years later, an
explanation for this observation was given by Albert Einstein, who
traced it to the random collisions between the grains and the water
molecules \cite{einstein}. These collisions cause the Brownian
particle to exhibit a ``random walk'' in space - a phenomenon also
known as single particle diffusion. Brownian motion can be described
by the diffusion equation for the particle's probability density
function (PDF), which in one-dimension (1D) reads
\begin{equation}
  \frac{\partial P(x,t)}{\partial t}=D\frac{\partial^2 P(x,t)}{\partial
    x^2},
  \label{eq:diffsimple}
\end{equation}
where $x$ and $t$ denote, respectively, coordinate and time, and $D$
is the diffusion coefficient. Assuming a Dirac delta-function initial
condition $P(x,0)=\delta(x)$, the solution of Eq.~(\ref{eq:diffsimple})
is given by the normal distribution $P(x,t)=\left(4\pi
Dt\right)^{-0.5}\exp\left(-x^2/4Dt\right)$. The mean displacement of
the Brownian particle $\left\langle \Delta x\right\rangle=0$, while
the mean squared displacement (MSD) grows linearly with time:
$\left\langle\Delta x^2\right\rangle=2Dt$.

An alternative route for describing the motion of a Brownian particle
is the Langevin equation \cite{langevin}
\begin{equation}
  m\frac{dv}{dt}=-\alpha v+\beta(t),
  \label{eq:langevinsimple}
\end{equation}
where $m$ and $v$ denote, respectively, the mass and velocity of the
particle. In this description, the impact of the random collisions on
the Brownian particle is realized through the action of two forces,
represented by the terms on the r.h.s.~of the equation. The first is a
friction force representing the statistical average of the collision
forces, while the second is a random Gaussian white noise accounting
for the force distribution around the mean value. Since the magnitude
of the collision forces depends on the characteristic thermal velocity
of the molecules of the embedding fluid, the friction coefficient
$\alpha$ in Langevin's equation (\ref{eq:langevinsimple}) must depend
on the temperature, $T$. It should also be related to the diffusion
coefficient, $D$, appearing in Eq.~(\ref{eq:diffsimple}). The relation,
$\alpha=k_BT/D$ (where $k_B$ is Boltzmann's constant), is known as
Einstein's relation, which is closely related to the more general
fluctuation-dissipation theorem \cite{kubo}. In order for the latter
to be satisfied, one must also assume that the Gaussian white noise
term in Langevin's equation has the following statistical properties
\cite{parisi}: $\langle \beta(t)\rangle =0$, and $\langle
\beta(t)\beta(t^{\prime})\rangle=2\alpha k_BT\delta(t-t^{\prime})$,
where $\langle\cdots\rangle$ denotes average over all possible
realizations of the noise force $\beta(t)$.

Two comments regarding Langevin's equation (\ref{eq:langevinsimple})
should be made. First, the equation describes Brownian diffusive
dynamics only at large time scales. On short time scales, Langevin
dynamics is ballistic (inertial). The crossover between the ballistic
and diffusive regimes occurs at $\tau\sim m/\alpha$. Second, the
equation neglects the influence of the motion of the Brownian particle
on the embedding fluid. The fluid acts as an ideal heat bath whose
properties remain unaffected by the presence of the moving Brownian
particle. This latter assumption is justified when the number of fluid
molecules is macroscopically large, and when the momentum and energy
are locally transferred to the bulk fluid much faster than any other
relevant time scale of the dynamics.

The difference between the diffusion equation approach to Brownian
dynamics and the Langevin equation formalism becomes more significant
when one deals with diffusion in a medium with a position-dependent
friction coefficient $\alpha(x)$. These types of dynamics are often
associated with the It\^{o}-Stratonovich dilemma \cite{mannella}. In
this paper we study specific examples of such dynamics occuring in
systems with power-law diffusion coefficients $ D(x)\sim |x|^c$. Model
systems with spatially-dependent diffusivity have been receiving
renewed interest recently due to their relevance to single particle
experiments involving femto-Newton force measurements
\cite{lancon,volpe}. The dilemma itself is not the main topic of this
paper, and we refer the reader to textbooks on stochastic dynamics,
e.g.~\cite{coffey,kampen}, for more details. Here we summarize only
the highlights relevant to this work:

1. The generalization of Eq.~(\ref{eq:diffsimple}) corresponding to
dynamics of Brownian particles, at constant temperature, in (1D)
systems with spatially-dependent friction coefficients, is~\cite{lau}
\begin{equation}
\frac {\partial P(x,t)}{\partial t} =\frac {\partial}{\partial
  x}\left(D\left(x\right) \frac{\partial P(x,t)}{\partial x}\right).
\label{eq:diffusion}
\end{equation}
This is Fick's second law $\partial_t P=-\partial_x J$, with the flux
$J(x,t)=-D(x)\partial_x P(x,t)$.

2. The corresponding Langevin equation reads \cite{gjf3}
\begin{equation}
m\frac {dv}{dt}=-\alpha(x) v+\beta\left(x\left(t\right)\right),
\label{eq:langevin}
\end{equation}
with $\alpha(x)=k_BT/D(x)$, which is a natural generalization of
Einstein's relation \cite{gjf4}.

3. A dilemma arises when the Langevin equation (\ref{eq:langevin}) is
integrated over time in order to calculate the trajectory of the
particle \cite{domb}. Since the particle moves during the
infinitesimal time step $dt$, the value of $\alpha(x)$ also changes,
and the Langevin equation of motion must be supplemented with a
convention (rule) for choosing the value of $\alpha(x)$. The name
``It\^{o}-Stratonovich dilemma'' assigned to the ambiguity about the
choice of interpretation is after the two most commonly used
conventions - the one of It\^{o} which uses the value of $\alpha$ at
the beginning of the time step, and the one of Stratonovich which
takes the average of the friction function at the initial and the end
points.

4. In the overdamped limit, i.e., when the inertial term on the
l.h.s.~of Eq.~(\ref{eq:langevin}) is set identically to zero,
different conventions lead to trajectories with different statistical
properties, even for $dt\rightarrow 0$ \cite{risken}. For Brownian
dynamics at constant temperature, the correct convention that
generates the PDF solving Eq.~(\ref{eq:diffusion}) is neither
It\^{o}'s nor that of Stratonovich, but rather H\"{a}nggi's
interpretation (also known as the ``isothermal'' convention)
\cite{hanggi,hanggiremark} which uses $\alpha$ at the end of the time
step \cite{lau,gjf3}.

5. In the case of underdamped Langevin dynamics [i.e.,
  Eq.~(\ref{eq:langevin}) with the l.h.s.~{\em not}\/ assumed to be
  vanishingly small], all (reasonable) conventions converge to the
correct PDF in the limit $dt\rightarrow 0$. This difference between
underdamped dynamics and its overdamped limiting case (see item 4
above) stems from fact that in the latter, the velocity is physically
ill-defined (since it is proportional to the white noise $\beta$),
while in the former, it remains finite and follows the equilibrium
Maxwell-Boltzmann distribution. Formally (mathematically) speaking,
there is no dilemma in the second-order (in $x$) equation
(\ref{eq:langevin}). However, the rate of convergence of numerical
simulation-results toward the theoretical PDF [i.e., the solution of
  Eq.~(\ref{eq:diffusion})] greatly depends on the chosen convention
and the numerical integrator. This issue has considerable practical
importance in numerical simulations where the time step $dt$ is not
infinitesimal. The results in the work are based on Langevin dynamics
simulations employing the G-JF integrator \cite{gjf1,gjf2} with a
newly proposed ``inertial'' convention \cite{gjf3,gjf4} (see details
in section \ref{subsec:ldynamics}). This combination produces
excellent results even for relatively large integration time steps.

With the above in mind, we now turn to examine the behavior of a
Brownian particle moving, at constant temperature, in a 1D system with
a power-law diffusion function $D(x)=D_0\left | x/l \right |^c$. In
the following section we calculate the PDF of the particle by solving
the diffusion equation (\ref{eq:diffusion}), and by numerically
integrating the Langevin equation of motion (\ref{eq:langevin}). As we
will observe, these two approaches do not necessarily yield similar
results.

\section{Heterogeneous media with power law friction function}

\subsection{Fick's second law}
\label{subsec:fickslaw}

For $D(x)=D_0\left | x/l \right |^c$, the solution of
Eq.~(\ref{eq:diffusion}) is
\begin{equation}
P(x,t)=\frac{\left[\left(2-c\right)^cD_0t\right]^{1/(c-2)}}{2\Gamma
  \left (\frac{1}{2-c} \right
  )}\exp\left[\frac{-|x|^{2-c}}{(2-c)^2(D_0t)}\right],
\label{eq:solution}
\end{equation}
where $\Gamma$ is the Gamma function, and for brevity we set
$l=1$. This solution satisfies the condition that the particle's
motion starts at the origin: $P(x,0)=\delta(x)$. From the requirement
that $P(x,t)$ vanishes for $x \rightarrow \pm \infty $, which is
necessary (but not sufficient) to ensure that $\int_{-\infty}
^{\infty}P(x,t)dx=1$, we infer that the solution can be physical only
for $c<2$.  From symmetry considerations, the ensemble average $\left
\langle x \right \rangle=0$, while the MSD,
\begin{equation}
 \left \langle x^2 \right \rangle=\int_{-\infty} ^{\infty}\!\!\!
 x^2P(x,t)dx=\frac{\Gamma\left ( \frac{3}{2-c}
   \right )}{\Gamma\left ( \frac{1}{2-c} \right )}
 \left[\left(2-c\right)^2D_0t\right]^{2/(2-c)}.
\label{eq:xsquared}
\end{equation}
Thus, for $c<0$ we observe subdiffusion, and for $0<c<1$ we find
superdiffusion. For $c=0$ we have $\left\langle
x^2\right\rangle=2D_0t$, i.e., normal diffusion, and for $c=1$ the
particle's motion is ballistic. For $c>1$, Eq.~(\ref{eq:xsquared})
predicts dynamics which are faster than ballistic (for instance,
$c=1.5$ corresponds to dynamics at costant acceleration). This is an
unphysical result, and in what follows we demonstrate that for any
$c\geq 1$ the motion remains ballistic.

\subsection{Langevin Dynamics Simulations}
\label{subsec:ldynamics}

The PDF can be obtained from an ensemble of trajectories of particles
starting at the origin, $x^0=0$, with initial velocities, $v^0$, drawn
from an equilibrium Maxwell-Boltzmann distribution. The trajectories
are computed by numerically integrating Langevin's equation of motion
(\ref{eq:langevin}). Denoting, respectively, by $x^n$ and $v^n$ the
position and velocity of a particle at time $t_n$, the integration is
conducted using the G-JF algorithm that advances the system by one time
step to $t_{n+1}=t_n+dt$, using the following set of discrete-time
equations \cite{gjf1,gjf2}:
\begin{eqnarray}
  x^{n+1}&=&x^n+bdtv^n+\frac{bdt^2}{2m}f^n+\frac{bdt}{2m}\beta^{n+1}
  \label{eq:gjfx}\\
  v^{n+1}&=&av^n+\frac{dt}{2m}\left(af^n+f^{n+1}\right)+\frac{b}{m}\beta^{n+1},
  \label{eq:gjfv}
\end{eqnarray}
where $f^n=f\left(x^n\right)$ is the {\em deterministic}\/ force
acting on the particle, $\beta^{n+1}$ is a Gaussian random number with
\begin{eqnarray}
\left\langle \beta^n\right\rangle=0\ ;\  \left\langle
\beta^n\beta^l\right\rangle=2\alpha k_BT dt\delta_{n,l},
\label{eq:noise}
\end{eqnarray}
and the damping coefficients of the algorithm are
\begin{eqnarray}
  b=\left[1+\left(\alpha dt/2m\right)\right]^{-1}\ ;
  \ a=\left[1-\left(\alpha dt/2m\right)\right]b.
  \label{eq:constants}
\end{eqnarray}
We set $f^n=0$ since we consider the case when the particle
experiences no forces other than random collisions with the fluid
molecules.

Since the friction coefficient varies in space, the above equations
(\ref{eq:gjfx}) and (\ref{eq:gjfv}) must be complemented with a
convention for choosing the value of $\alpha$ to be used in
Eqs.(\ref{eq:noise}) and (\ref{eq:constants}) at each time step. Here,
we use the recently proposed inertial convention that assigns to
$\alpha$ the value of the spatial average of the friction function
along the inertial trajectory from $x^n$ to
$\tilde{x}^{n+1}=x^n+v^ndt$ \cite{gjf3,gjf4}
\begin{equation}
\frac{\int_{x^n}^{\tilde{x}^{n+1}} \alpha (x)
  dx}{\tilde{x}^{n+1}-x^n}=\frac{A(\tilde{x}^{n+1})-A(x^n)}{\tilde{x}^{n+1}-x^n},
\label{eq:avgalpha}
\end{equation}
where $A(x)$ is the primitive function of $\alpha(x)$. We have
previously demonstrated that the combination of the G-JF algorithm
with the inertial convention produces excellent agreement between the
computed and theoretical PDFs, even for relatively large intergration
time steps.

\begin{figure}[t]
\centering\includegraphics[width=0.45\textwidth]{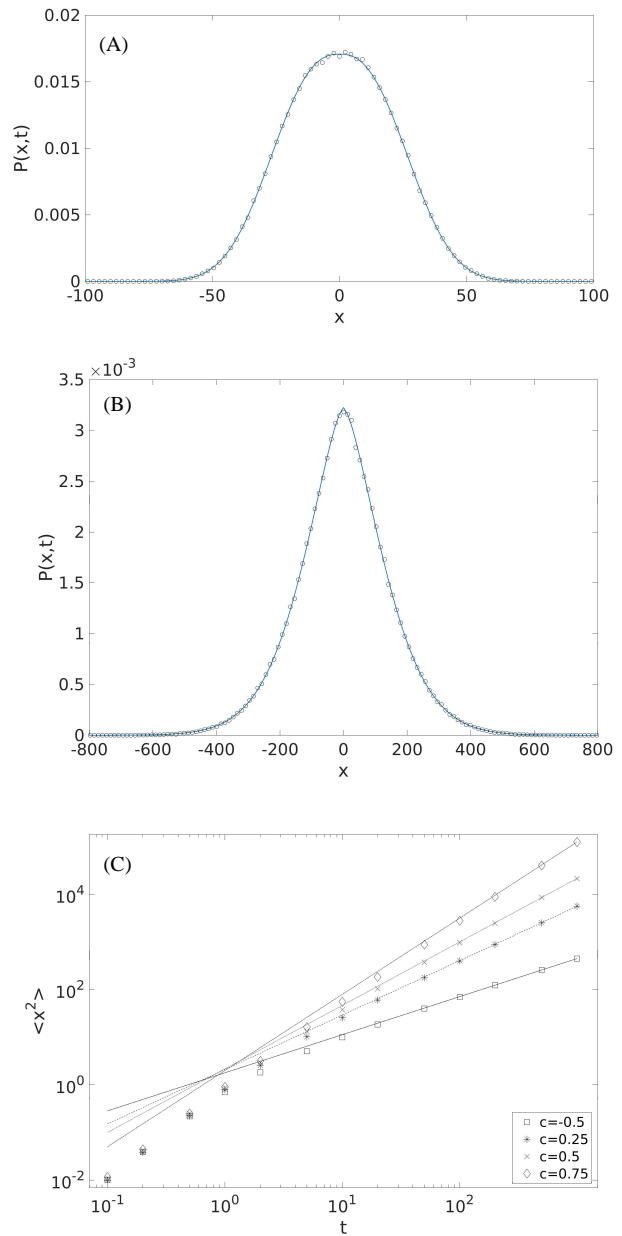} 
\caption{{\bf (A)} The PDF at $t=1000$ obtained from the numerical
  integration (open circles) vs.~the PDF given by
  Eq.~(\ref{eq:solution}) (solid line), which solves Fick's diffusion
  equation for $c=-0.5$. {\bf (B)} Same as {\bf (A)}, for $c=0.5$.
  {\bf (C)} The MSD $\langle x^2 \rangle$ of the particle as a
  function of time for $c=-0.5,0.25,0.5,0.75$ (markers), vs. the
  expected $\langle x^2 \rangle$ for $c=-0.5,0.25,0.5,0.75$ according
  to Eq.~(\ref{eq:xsquared}) (lines).}
\label{fig:fig1}
\end{figure}

Fig.~\ref{fig:fig1} depicts our results for the PDF for systems with a
power law friction function
$\alpha(x)=k_BT/D(x)=(k_BT/D_0)|x/l|^{-c}$, for $c=-0.5$ (A) and
$c=+0.5$ (B). For convenience, we set $m=1$, $D_0=1$, $k_BT=1$, and
$l=1$. The results have been obtained from simulations of
$2.5\times10^5$ trajectories with integration time step
$dt=10^{-3}$. The open circles in Figs.~1(A) and (B) represent our
numerical results for the PDF at $t=1000$ for $c=-0.5$ and $c=0.5$,
respectively. The numerical results exhibit excellent agreements with
the corresponding analytical predictions of Eq. (\ref{eq:solution}),
which are plotted with the solid curves.

The symbols in Fig.~\ref{fig:fig1}(C) represent the numerical results
for the MSD for $c=-0.5,0.25,0.5,0.75$. We observe that the power-law
behavior $\langle x^2 \rangle\sim t^{\frac{2}{2-c}}$ [see
  Eq.~(\ref{eq:xsquared})], which is depicted by the lines in the
figure, is indeed recovered at large times.  The same power-law
(\ref{eq:xsquared}) was previously derived in ref.~\cite{metzler},
where instead of Fick's law (\ref{eq:diffusion}), a different
diffusion equation
$\partial_tP=\partial_x\left[\sqrt{D\left(x\right)}\,
  \partial_x\left(\sqrt{D\left(x\right)}P\right)\right]$ was
considered. The latter form of the diffusion equation corresponds to
the Stratonovich interpretation of the overdamped Langevin
equation. The reader is reminded (see section \ref{sec:intro}) that
for overdamped Langevin dynamics, different conventions lead to
different PDFs. Indeed, although both equations yield the same
power-law for the MSD, the PDFs solving these equations look markedly
different. Specifically, the PDFs of the Statonovich diffusion
equation diverge at the origin for $c>0$, and assume a a bimodal form
for $c<0$, with a vanishing value at the origin \cite{metzler}. In
contrast, the PDFs of Fick's law of diffusion (which corresponds to
H\"{a}nggi's interpretation) attain a maximum at the origin.  Our
Langevin dynamics simulations, which at large times reproduce PDFs
that agree with Eq.~(\ref{eq:solution}), serve as yet another
demonstration for the appropriateness of H\"{a}nggi interpretation and
Fick's second law for diffusion at constant temperature. This is
because the simulations follow the underdamped (inertial) Langevin
dynamics of the particle. As noted above, for the inertial Langevin
equation, all interpretations converge to the correct solution in the
limit of small integration time steps.

\section{Ballistic Motion}

\subsection{The case $c\geq 1$}
\label{subsec:clarger1}

Integrating Eq.~(\ref{eq:langevin}) from the initial time to $t$, and
taking the ensemble average over all noise realizations, yields the
following relationship
\begin{eqnarray} 
\left\langle m\Delta v\right\rangle=-\left\langle \Delta
A\left(x\right)\right\rangle
\label{eq:relation1}
\end{eqnarray}
between the momentum change (force impulse) and displacement of the
particle. Eq.~(\ref{eq:relation1}), which was previously derived in
ref.~\cite{gjf4}, involves $A(x)$ - the primitive function of
$\alpha(x)$ [see Eq.~(\ref {eq:avgalpha})]. This implies that
$\alpha(x)$ must be an integrable function. For $\alpha(x)\sim x^{-c}$
with $c\geq 1$, the friction function is non-integrable at $x=0$. This
feature makes it impossible to start the simulations when the particle
is at the origin due to the inability to define the friction
coefficient for the initial step. If the particle is placed on one
side of the system, it will never cross to the other side. This is
because no matter how close the particle approaches to the origin, its
ballistic distance (the characteristic distance that it travels before
changing its direction) will always be shorter than the distance to
the origin. In other words, for $c\geq 1$, the dissipation near the
origin diverges so rapidly, that the singularity acts like a wall that
stops the particle and bounces it back. This scenario, however, is
unphysical, and it stems from the unphysical nature of Langevin's
equation which only considers the influence of the medium on the
particle but ignores the impact of the particle on the medium. From
momentum conservation we know that any change in the momentum of the
Brownian particle must be countered by an opposite change in the
momentum of the molecules of the medium. This implies that when the
particle is reflected from the origin, it exerts a force on the
friction singularity, and this force will cause changes in the medium
that would not allow the singularity to be long-lived.

Apart from the divergence of $A(x)$ at the origin, it is also
interesting to consider the ramifications of the rapid drop in
$\alpha(x)$ in the limits $x \rightarrow \pm \infty$. For $c>1$, the
integral over $\alpha(x)$ from $x_0>0$ ($x_0<0$) to $+\infty$
($-\infty$) is finite, implying that the particle's ballistic distance
may diverge. This can be inferred from Eq.~(\ref{eq:relation1}), which
suggests that it is unlikely for a particle to change its direction of
motion, if it reaches $x_0$ with velocity
$v_0>\left[A\left(\infty\right)- A\left(x_0\right)\right]/m$. In other
words - as the particle travels further away from the origin, it
experiences a vanishingly small friction force and, therefore, its
motion would ultimately become ballistic. The crossover from diffusive
to ballistic dynamics is further explored in the following section.

\subsection{Crossover to ballistic motion}
\label{subsec:crossover}

We now consider dynamics in a one-dimensional system with the
spatially-dependent diffusion coefficient
\begin{equation}
 D(x)=D_0\left[1+\left(\frac{x}{l}\right)^2\right].
\label{eq:dball}
\end{equation}
For this choice, $D(x)\sim x^c$ with $c=2$ for $x/l\gg 1$; but unlike
the power-law form discussed in section \ref{subsec:clarger1} above,
the friction coefficient, $\alpha(x)=k_BT/D(x)$, does not diverge at
the origin.  A special reason for choosing the specific form
Eq.~(\ref{eq:dball}) is that it has been given in ref.~\cite{lau} as an
example of a spatially-dependent diffusion coefficient causing
increasing acceleration. This result is obtained by multiplying by
$x^2$ both sides of the diffusion equation (for $D_0=1$ and $l=1$)
\begin{eqnarray}
\frac {\partial P(x,t)}{\partial t} =\frac {\partial}{\partial
  x}\left[\left(1+x^2\right) \frac{\partial P(x,t)}{\partial
    x}\right],
\label{eq:diffusion2}
\end{eqnarray}
and integrating with respect to $x$, which yields the equation
\begin{eqnarray}
  \frac {\partial \left\langle x^2\right\rangle}{\partial t} =
  2+6\left\langle x^2\right\rangle,
\label{eq:diffusion3}
\end{eqnarray}
which has the solution
\begin{equation}
\left\langle
x^2\right\rangle=\left[\exp\left(6t\right)-1\right]/3.
\label{eq:wrongmsd}
\end{equation}
However, the prediction of Eq.~(\ref{eq:wrongmsd}) that the MSD grows
exponentially with $t$ is unphysical since it implies the emergence of
two opposite currents of particles with ever-increasing
velocities. One should wonder about the energy source of the
exponential growth in the kinetic energy of the Brownian particles.
The particles are immersed in a medium of uniform temperature serving
as a heat bath, and experience no force other than random collisions
with the molecules of the medium. It is impossible that through random
collisions, the Brownian particles would consistently gain energy
allowing them to reach exponentially large speeds, especially at large
distances where the friction coefficient vanishes, which means that
the rate of collisions with the heat bath becomes increasingly small.

\begin{figure}[t]
\centering\includegraphics[width=0.45\textwidth]{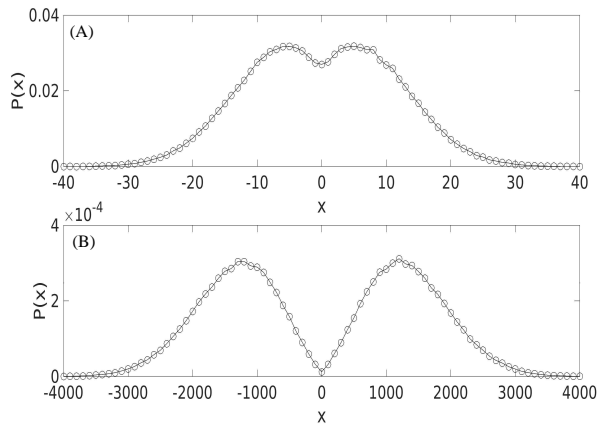} 
\caption{The PDF at $t=10$ {\bf (A)} and $t=1000$ {\bf (B)} of a
  Brownian particle starting at the origin and moving in a medium
  where $\alpha(x)=1/(1+x^2)$.}
\label{fig:fig2}
\end{figure}

The erroneous Eq.~(\ref{eq:wrongmsd}) is derived from the diffusion
equation (\ref{eq:diffusion2}). The latter, however, does not
correctly depict the dynamics of the particles in the system because
is applies only to time-scales much larger than the ballistic time of
the motion. As noted at the end of section \ref{subsec:clarger1}, the
ballistic distance diverges when the friction function drops faster
than $x^{-1}$ at large distances. When this occurs, the velocity of
the particle saturates to some finite value, and the dynamics becomes
ballistic. In other words, the ballistic time diverges, and the
dynamics never reaches the diffusive regime of Eq.~(\ref{eq:diffusion2}).

\begin{figure}[t]
\centering\includegraphics[width=0.48\textwidth]{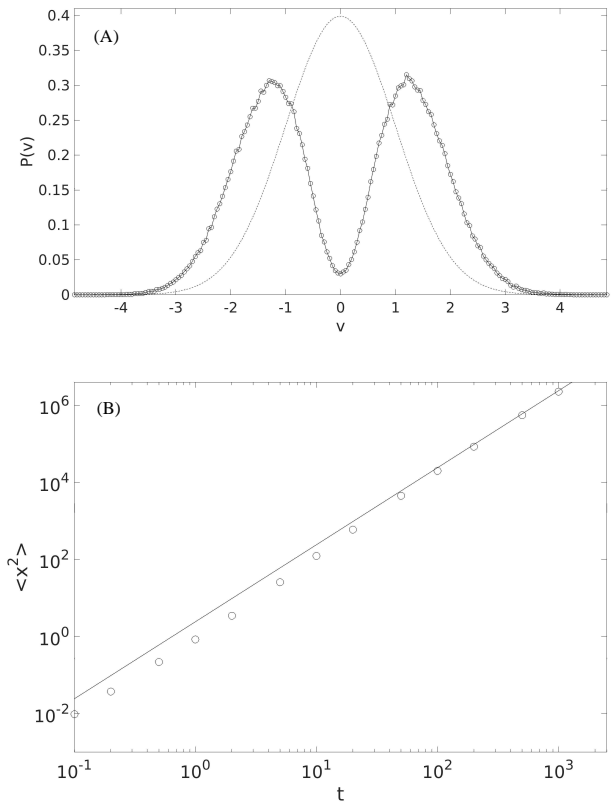} 
\caption{{\bf (A)} Circles - The bimodal VPDF at $t=1000$. The dashed
  line depicts the initial Gaussian equilibrium VPDF. {\bf (B)} The
  computed MSD of the particle (circles) vs.~the asymptotic power-law
  form $\left\langle x^2\right\rangle=2.41 t^2$ (solid line).}
\label{fig:fig3}
\end{figure}

In contrast to the diffusion equation (\ref{eq:diffusion}), the
Langevin equation (\ref{eq:langevin}) applies to both the ballistic
and diffusive regimes. Fig~\ref{fig:fig2} presents our results for the
PDF of the particles at $t=10$ (A) and $t=1000$ (B). The results,
which are based on numerical intergation of $3.25\times10^5$
trajectories starting at the origin (with velocities drawn from the
standard Gaussian equilibrium distribution), demonstrate that as the
time increases, the PDF becomes increasingly bimodal. This indicates
the emergence of two opposite particle currents propagating away from
the origin.  Fig.~\ref{fig:fig3}(A) shows the velocity probability
distribution function (VPDF) at $t=1000$ (solid circles), which is
also bimodal and, thus, does not coincide with the initial equilibrium
distribution (depicted by the dashed line in the figure). The VPDF at
$t=2000$ (not shown) is essentially identical to the VPDF in
Fig.~\ref{fig:fig3}(A), which proves this VPDF represents the steady
state of the velocity distribution. From the steady state VPDF, we
find that the steady state squared velocity $\left\langle
v^2\right\rangle\simeq 2.41$ (in units of $k_BT/m$) and, therefore, at
large times the position MSD $\left\langle
x^2\right\rangle=\left\langle v^2\right\rangle t^2=2.41 t^2$. This
result, which is fully corroborated by the numerical data in
Fig.~\ref{fig:fig3}(B), demonstrates that the particles end up moving
inertially with velocities drawn from the steady state VPDF.

\subsection{The fluctuation-dissipation relationship}
\label{subsec:fdr}

Integrating Eq.~(\ref{eq:langevin}) with respect to time, squaring the
equation, and taking the ensemble average over noise realizations,
yields the generalized form of the fluctuation-dissipation
relationship for systems with spatially varying friction~\cite{gjf4},
which reads
\begin{equation}
\left\langle \left(m\Delta v\right)^2+2m\Delta v\Delta A+\left(\Delta
A\right)^2 \right \rangle =\int_{0}^t
2\left\langle\alpha\left(t^{\prime}\right)\right \rangle k_B
Tdt^{\prime}.
\label{eq:fdr}
\end{equation}
If at large times the dynamics enters the diffusive regime, the first
two terms on the l.h.s.~become negligible compared to the third
one. Moreover, for a constant $\alpha$, the third term on the
l.h.s.~is equal to $\alpha^2\left\langle \left(\Delta
x\right)^2\right\rangle$, while the integral on the r.h.s.~gives
$2\alpha k_BTt$. Thus, for constant friction, expression
(\ref{eq:fdr}) reduces (at large times) to the well-known form of the
fluctuation-dissipation relationship $\left\langle \left(\Delta
x\right)^2\right\rangle =2(k_BT/\alpha)t=2Dt$.

\begin{figure}[t]
\centering\includegraphics[width=0.45\textwidth]{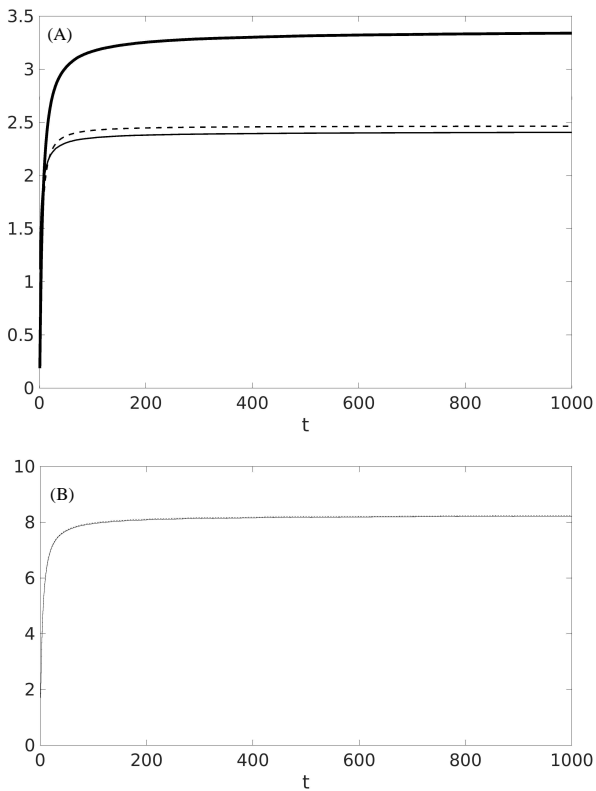} 
\caption{{\bf (A)} The ensemble averages of $(m\Delta v)^2$ [first
    term on the l.h.s.~of Eq.~(\ref{eq:fdr}) - thin solid line],
  $2m\Delta v\Delta A$ (second term - thick solid line), and $(\Delta
  A)^2$ (third term - dashed line), as a function of time $t$. {\bf
    (B)} The sum of the three averages shown in (A) (dashed line)
  vs.~the average of $2\alpha k_BT$ integrated over time [r.h.s.~of
    Eq.~(\ref{eq:fdr}) - solid line], as a function of $t$.}
\label{fig:fig4}
\end{figure}

For the dynamics discussed in section \ref{subsec:crossover}, the
motion is not diffusive, but rather becomes ballistic at large
times. However, relationship (\ref{eq:fdr}) holds for any time $t$,
regardless of the character of the dynamics. This is nicely
demonstrated in Fig.~\ref{fig:fig4}(A), where we plot the ensemble
averages of the three terms on the l.h.s.~of Eq.~(\ref{eq:fdr}). As can
be seen, all three terms grow rapidly at short times, which include
the very initial ballistic segment, and the following interval of
diffusive motion. At $t\gtrsim 100$, all three terms saturate, which
indicates the crossover from diffusive to ballistic motion. In (B) we
plot the sum of the averages of these three terms (dashed line)
vs.~the value of the r.h.s.~of Eq.~(\ref{eq:fdr}), which is the
ensemble average of $\alpha\left(x\left(t^{\prime}\right)\right)$
integrated from the beginning of the dynamics until time $t$ (solid
line). The lines overlap each other (the relative difference between
them is smaller than $1\%$), which demonstrates that the equality
between the two sides of Eq.~(\ref{eq:fdr}) holds at all times.

\section{Summary and Discussion}

In this work we used computer simulations to study the Langevin
dynamics of Brownian particles in a 1D system with a friction
coefficient that varies as a power-law of the distance from the origin
$\alpha(x)\sim |x|^{-c}$. It has been demonstrated that for $c<1$, the
particle diffuses anomalously with MSD $\left\langle
x^2\right\rangle\sim t^{2/(2-c)}$. This result can be also derived by
solving the corresponding diffusion equation (\ref{eq:diffusion}) with
diffusion coefficient $D(x)=k_BT/\alpha(x)$.

The diffusion equation can be formally solved for $c<2$. For $1<c<2$,
the solution incorrectly predicts that the MSD grows faster than the
MSD of ballistic motion. This result stems from the diffusion
equation, which cannot be physically justified for time scales smaller
than the ballistic time of the motion. For constant friction
coefficient $\alpha$, a crossover from ballistic to diffusive motion
occurs on time scales $\tau\geq m/\alpha$. In the case when
$\alpha(x)\sim |x|^{-c}$ with $c>1$, the friction vanishes rapidly at
large distance and an opposite crossover, from diffusive back to
ballistic motion, takes place. When this happens, the diffusion
equation can no longer be used if the ballistic time diverges, in
which case the motion remains ballistic (as the particle escapes to
infinity).

In the example discussed in section \ref{subsec:crossover}, the
divergence of the friction at the origin is removed, while at large
distances $\alpha(x)\sim |x|^{-2}$. Even if the friction never
vanishes completely, it drops at such a fast rate that it quickly
becomes irrelevant. Thus, this example resembles dynamics of a
Brownian particle in a finite fluid drop held at constant temperature
$T$. When the particle reaches the surface of the drop, it escapes,
and its velocity no longer changes. At the moment of escape the
particle has to have a velocity component directed outwards from the
drop and, therefore, the velocity distribution function outside of the
drop differs from the equilibrium Gaussian equilibrium velocity
distribution at temperature $T$ [see
  Fig.~\ref{fig:fig3}(A)]. Noticeably, the mean kinetic energy of the
escaping particle is larger than the corresponding equilibrium value,
$dk_BT/2$ (where $d$ is the dimentionality of the system). The fact
that, on average, the escaping particle takes away an amount of
kinetic energy larger than the equilibrium value implies that the
molecules of the fluid drop are left with an average kinetic energy
smaller than the equilibrium value. The drop cools down slightly, and
in order to maintain the temperature at $T$, it must be connected to a
true heat reservoir that would supply the missing energy. This
consideration is missing in the framework of Langevin's equation that
completely neglects the influences of the Brownian particle on the
surrounding medium.

Finally, we note that, within the framework of Langevin dynamics, a
generalized form of the fluctuation-dissipation relation has been
previously derived [Eq.~(\ref{eq:fdr})]. This form holds for dynamics
in media with spatially varying friction, at all times (i.e., both
within the ballistic and diffusive regimes of the dynamics). If
$\alpha(x)$ is bound between two positive values, the motion at large
times becomes diffusive. In such a case, the l.h.s.~of
Eq.~(\ref{eq:fdr}) becomes dominated by the third term, and both sides
of the equation grow linearly with $t$.  Anomalous diffusion is
observed when at large distances $\alpha(x)\sim |x|^{-c}$ with
$c<1$. In this case, the third term still dominates the other two
terms on the l.h.s.; however, the large time behavior scales like
$x^{2(1-c)}\sim t^{2(1-c)/(2-c)}$. For $c\geq 1$, the motion becomes
ballistic at large times. In this case, all terms on the l.h.s.~are
equally important and, like the r.h.s.~of the equation, relax to
constant asymptotic values.


\begin{thebibliography}{99}

\bibitem{brown} R. Brown, Philos. Mag. {\bf 4}, 161 (1828).

\bibitem{einstein} A. Einstein, Ann. Phys. (Leipzig) {\bf 17}, 549
  (1905).
  
\bibitem{langevin} P. Langevin, C. R. Acad. Sci. (Paris) {\bf 146},
  530 (1908).

\bibitem{kubo} R. Kubo, Rep. Prog. Phys. {\bf 29}, 255 (1966).

\bibitem{parisi} G. Parisi, {\em Statistical Field Theory}\/ (Addison
  Wesley, Menlo Park, 1988).

\bibitem{mannella} R. Mannella and V. P. E. McClintock, Fluct. Noise
  Lett. {\bf 11}, 1240010 (2012).

 \bibitem{lancon} P. Lan\c{c}on, G. Batrouni, L. Lobry, and N. Ostrowsky,
  Europhys. Lett. {\bf 54}, 28 (2001).

 \bibitem{volpe} G. Volpe, L. Helden, T. Brettschneider, J. Wehr, and
   C. Bechinger, Phys. Rev. Lett.  {\bf 104} , 170602 (2010).
 
\bibitem{coffey} W. T. Coffey, Y. P. Kalmyfov, and J. T. Waldron, {\em
  The Langevin Equation: with application in Physics, Chemistry, and
  Electrical Engineering}\/ (World Scientific, London, 1996).

\bibitem{kampen} N. G. van Kampen, {\em Stochastic Processes in
  Physics and Chemistry}\/ (North-Hollans, Amsterdam, 1981).

\bibitem{lau} A. W. C. Lau and T. C. Lubensky, Phys. Rev. E {\bf 76},
011123 (2007).

\bibitem{gjf3} O. Farago and N. Gr\o nbech-Jensen, Phys. Rev. E {\bf
89}, 013301 (2014).
 
\bibitem{gjf4} O. Farago and N. Gr\o nbech-Jensen, J. Stat. Phys. {\bf
156}, 1093 (2014).

\bibitem{domb} J. L. Domb, Ann. Math. {\bf 43}, 351 (1942).

\bibitem{risken} H. Risken, {The Focker-Planck Equation}\/
  (Springer-Verlag, Berlin, 1988).

\bibitem{hanggi} P. H\"{a}nggi, Helv. Phys. Acta {\bf 51} , 183
  (1978).

\bibitem{hanggiremark} Stochastic differential equations can be used
  to describe different processes in diverse fields, such as biology,
  economy, neuroscience, geophysics, and more. For many applications,
  other interpretations (e.g., It\^{o} or Stratonovich) may be
  appropriate. Here, we consider the physical problem of a particle's
  dynamics in isothermal systems, for which the relevant
  interpretation is the one of H\"{a}nggi. When the particle moves in
  a potential energy field $U(x)$, a force term $f(x)=-dU/dx$ should
  be added to the r.h.s.~of Eq.~(\ref{eq:langevin}). At large times,
  one expects to approach the equilibrium Boltzmann distribution
  function $P(x)\sim \exp\left[-U\left(x\right)/k_BT\right]$, which
  does not depend on the diffusivity function $D(x)$. This is achieved
  with the H\"{a}nggi conventions, while other interpretations require
  the addition of a spurious drift term to the Langevin equation (see
  \cite{lau,gjf3} for a more detailed discussion).
 
\bibitem{gjf1} N. Gr\o nbech-Jensen, and O. Farago. Mol. Phys. {\bf
111}, 983 (2013).

\bibitem{gjf2} N. Gr\o nbech-Jensen, N. R. Hayre, and O. Farago,
Comput. Phys. Commun. {\bf 185}, 524 (2014).


\bibitem{metzler} A. G. Cherstvy , A. V. Chechkin, and R. Metzler ,
  New J. Phys. {\bf 15}, 083039 (2013).


  
\end{thebibliography}
\end{document}